\documentclass{elsart}
\usepackage{amssymb,graphicx}

\begin{document}
\begin{frontmatter}
\title{Fast multilevel radiative transfer}
\author{Fr\'ed\'eric Paletou} \ead{fpaletou@ast.obs-mip.fr}
\author{\& Ludovick L\'eger}
\address[la2t]{Universit\'e Paul Sabatier, Toulouse 3, Observatoire
Midi-Pyr\'en\'ees, Laboratoire d'Astrophysique de Toulouse--Tarbes
(CNRS/UMR 5572), 14 ave. Edouard Belin, F-31400 Toulouse, France}

\begin{abstract}
The vast majority of recent advances in the field of numerical
radiative transfer relies on approximate operator methods better known
in astrophysics as Accelerated Lambda-Iteration
(ALI). A superior class of iterative schemes, in term of rates of
convergence, such as Gauss-Seidel and Successive Overrelaxation
methods were therefore quite naturally introduced in the field of
radiative transfer by Trujillo Bueno and Fabiani Bendicho~\cite{tf1}; it
was thoroughly described for the non--LTE two-level atom case. We
describe hereafter in details how such methods can be generalized when
dealing with non--LTE unpolarised radiation transfer with multilevel
atomic models, in monodimensional geometry.
\end{abstract}


\end{frontmatter}

\section{Introduction}

Our ability to deal with complex radiative transfer problems has
considerably improved during the last ten years. It is indeed still
worth an effort since important issues of astrophysical interest may
require to address a number of cases ranging from multi-dimensional
problems in various geometries (Auer and Paletou~\cite{lhafp}; Auer,
Fabiani Bendicho and Trujillo Bueno~\cite{lhaft} ; Fabiani Bendicho,
Trujillo Bueno and Auer~\cite{fta2}; van Noort, Hubeny and
Lanz~\cite{vanoort}; Gouttebroze~\cite{goutte}) to polarised radiation
transfer involving complex atomic models (Trujillo Bueno and Manso
Sainz~\cite{tm}; Trujillo Bueno~\cite{javier}), for instance.

Most of the recent developments in the field of numerical radiation
transfer are based on the Approximate (or Accelerated)
Lambda-Iteration (ALI) scheme which has been recently reviewed
in ~\cite{ivan}. ALI methods are based on operator splitting and, in a
seminal article, Olson, Auer and Buchler~\cite{oab} demonstrated the
merits of adopting an approximate operator which is nothing more than
the exact diagonal of the full operator $\Lambda$ allowing the
determination of the radiation field from a known source function (see
Mihalas~\cite{mihalas}; Rutten~\cite{rob}). More generally speaking
such an efficient iterative scheme is better known as the Jacobi's
method in numerical analysis (Young~\cite{young}).

However, even if superior -- in term of higher rates of convergence --
iterative schemes based on the Gauss-Seidel (GS) and the Successive
Overrelaxation (SOR) methods have been introduced since in the field
of numerical radiative transfer, they still deserved too little
attention by the astrophysical community. GS/SOR methods have been
described in details in a landmark article~\cite{tf1}; in particular,
the practical description of how to implement GS/SOR iterations within
a regular short characteristics formal solver is very meticulously
written and therefore extremely useful. However, their description of
the implementation of GS/SOR methods was restricted to the two-level
atom case in monodimensional (1D) geometry.

In a subsequent article~\cite{fta2}, although mentions to GS/SOR
iterative schemes generalized to the multilevel atom transfer problems
can be found, their implementation is far from being made explicit and
it is only described in very general terms; it is furthermore embedded
in another (very efficient though) numerical strategy based on
multi-grid methods~\cite{hackbusch,briggs}. Unfortunately, it appears
that the basic features of a GS/SOR-based formal solver in the frame
of the multilevel atom radiative transfer problem remain, so far, to be
explicited.

The present article aims indeed at filling the gap by providing all
the elements required for a successful implementation of {\em multilevel}
GS/SOR iterative schemes in 1D geometry.  We shall recall in \S 2 the
basic principles of GS/SOR iterative schemes in the case of a
two-level atom model. Then, in section \S 3 we shall describe step
by step how GS/SOR can be implemented for the case of multi-level atom
models, therefore extending the well-known multilevel-ALI method of
Rybicki and Hummer~\cite{mali1}. Finally, we shall present in \S 4 some
illustrative examples clearly demonstrating the performances of
multilevel GS/SOR iterative schemes.

\section{Gauss-Seidel and SOR iterative schemes basics}

Gauss-Seidel and SOR iterative scheme are particularly well adapted to
the {\em short characteristics} method (SC) for the formal solution of
the radiative transfer equation~\cite{lhafp,sc1,sc2}.

Using SC in 1D geometry indeed, the formal solution is obtained by
sweeping the grid say, first in direction $- \Omega$ ($\mu < 0$)
i.e. from the surface down to the bottom of the atmosphere, and then
in the opposite, upward direction $+ \Omega$ ($\mu > 0$) starting from
the bottom of the atmosphere up to its surface though. The specific
intensity $I_{\nu \Omega}$ is advanced step by step during each pass,
partially integrated over angles and frequencies during the downward
pass while, during the second (upward) pass, the mean intensity
$\bar{J}$ can be fully computed, completing therefore the formal
solution at each depth

\begin{equation}
\bar{J}_k = \Lambda [ S_k ] \, .
\end{equation}
Except at the boundaries where the illumination conditions are known a
priori, along each direction, the specific intensity at the inner grid
points is advanced depth after depth, and computed along the short
characteristics (allowing us to suppress any index related to the
angle variation of the specific intensity hereafter) according to

\begin{equation}
I_{o} = I_{u} e^{- \Delta \tau_{u}} + \Psi_{u} S_{u}+ \Psi_{o} S_{o}
+ \Psi_{d} S_{d}
\label{eq:sc}
\end{equation}
where the first part of the right-hand side of this expression
corresponds to the part transmitted from the ``upwind'' grid point $u$
down to the current point $o$ (see Fig. 1), and the three last term
result from the analytic integration of

\begin{equation}
{\Upsilon} = \int_{0}^{\Delta \tau_{u}} {S(\tau) e^{- \tau}} d \tau
\end{equation}
along the short characteristics going from $u$ to $o$. Indeed,
assuming that $S$ is quadratic in the optical depth $\tau$, it is easy
to show, using a simple Lagrange polynomial interpolation on the basis
of three consecutive grid points $u$, $o$ and $d$, that

\begin{equation}
{\Upsilon} = \Psi_{u} S_{u}+ \Psi_{o} S_{o} + \Psi_{d} S_{d} \, ,
\end{equation}
where the $\Psi$'s coefficients are defined\footnote{A different
expression was used in~\cite{lhafp}, Eqs. (8) to (10); note also that
there is a sign error in their coefficients $d_1$ and $d_2$. But more
important is to adopt the current expansion in terms of $\Psi$'s for
the efficient implementation of GS/SOR.} as

\begin{equation} 
\left\{
\begin{array}{l}
\Psi_{u} = \displaystyle{{ {w_2 + w_1 \Delta \tau_d} \over
{ \Delta \tau_{u} (\Delta \tau_{u} + \Delta \tau_{d})  } }} \\\\
\Psi_{o} = \displaystyle{ w_0 + {{w_1 (\Delta \tau_u - \Delta \tau_d) - w_2}
\over {\Delta \tau_{u} \Delta \tau_{d}} } }  \\\\
\Psi_{d} = \displaystyle{{ {w_2 - w_1 \Delta \tau_u} \over
{ \Delta \tau_{d} (\Delta \tau_{u} + \Delta \tau_{d})  } }}
\end{array}
\right. \, ,
\end{equation}
and where

\begin{equation} 
\left\{
\begin{array}{l}
w_{0} = 1 - e^{- \Delta \tau_u} \\
w_{1} = w_{0} - \Delta \tau_u e^{- \Delta \tau_u} \\
w_{2} = 2 w_{1} - \Delta \tau_{u}^{2} e^{- \Delta \tau_u}
\end{array}
\right. \, .
\end{equation}

In the \emph{two-level atom case}, the non--LTE line source function
$S$ is usually expressed as

\begin{equation}
S(\tau) = (1 - \varepsilon) \bar{J}(\tau) + \varepsilon B(\tau) \, ,
\end{equation}
where $\tau$ is the optical depth, $\varepsilon$ is the collisional
destruction probability measuring departures from LTE, $B$ the Planck
function and $\bar{J}$ is the usual mean intensity

\begin{equation}
\bar{J} = \int \displaystyle{{{d \Omega} \over {4 \pi}}}
  \int{\phi_{\nu} I_{\nu \Omega} d \nu} \, ,
\end{equation}
where we omitted the optical depth dependence for simplicity.

   \begin{figure}
   \centering
   \includegraphics[width=8cm,angle=0]{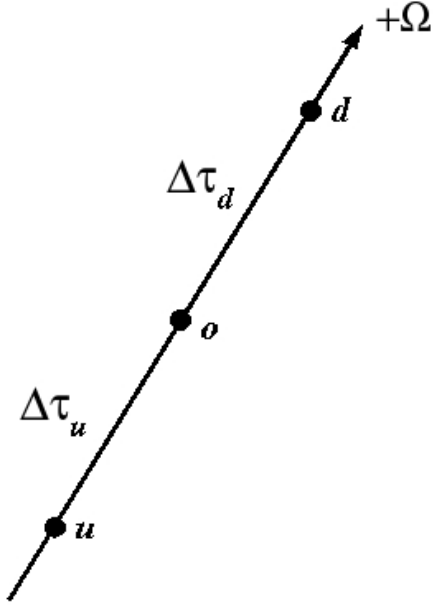}
      \caption{Three successive grid points along direction $+ \Omega$
              (upward) are used, assuming parabolic interpolation of
              the source function, in order to compute the specific
              intensity at current position $o$ along the short
              characteristics starting from $u$.  }
         \label{Fig1}
   \end{figure}

The Jacobi type iterative scheme introduced in numerical radiative
transfer by~\cite{oab} allows for the iterative determination
of the source function on the basis of increments such that

\begin{equation}
\Delta S_{k} = { 
{ (1 - \varepsilon) \bar{J}_{k}^{\mathrm{(old)}} + \varepsilon B_{k}
- S_{k}^{\mathrm{(old)}} } \over
{1 - (1 - \varepsilon) \Lambda_{kk} } } \, ,
\end{equation}
at each depth $\tau_{k}$, where the superscript (old) denotes
quantities already known from the previous iterative stage and where
$\Lambda_{kk}$ is a \emph{scalar} equal to the diagonal element of the
full operator $\Lambda$ at such a depth in the atmosphere.

Within a Gauss-Seidel iterative scheme the source function increments
now turn themselves into

\begin{equation}
\Delta S_{k}^{\mathrm{(GS)}} = { 
{ (1 - \varepsilon) \bar{J}_{k}^{\mathrm{(old\,and\,new)}} + \varepsilon B_{k}
- S_{k}^{\mathrm{(old)}} } \over
{1 - (1 - \varepsilon) \Lambda_{kk} } } \, ,
\end{equation}
where the somewhat enigmatic quantity
$\bar{J}_{k}^{\mathrm{(old\,and\,new)}}$ means that at the spatial
point $k$ the mean intensity has to be calculated via a formal
solution of the transfer equation using the ``new'' source function
values ${S}_{j}^{\mathrm{(new)}}$ already obtained at points $j=1,\,2,
\ldots, (k-1)$, and the ``old'' source function values
${S}_{j}^{\mathrm{(old)}}$ at points $j=k,\,(k+1), \ldots,
\mathrm{ND}$.

Achieving this was explained in every detail in~\cite{tf1} and we shall
not repeat here the algorithmics of the process. It is however
important to repeat that such a scheme is better implemented in the
frame of the short characteristics approach since it allows to avoid
any explicit matrix inversion.

Finally, going beyond Gauss-Seidel, SOR iterations can be implemented
where

\begin{equation}
\Delta S_{k}^{\mathrm{(SOR)}} = \omega \Delta S_{k}^{\mathrm{(GS)}} \, , 
\end{equation}
$\omega$ being, in the case of {\em overrelaxation}, a parameter such
that $1 < \omega < 2$, with an optimal value $\sim$ 1.5 for the two-level
atom case~\cite{tf1}.

However, in the non--LTE multilevel atom case, the radiation transfer
problem is somewhat more complicated since it is now required to solve
\emph{self-consistently} for the coupled set of $N_{\mathrm{trans}}$
radiative transfer equations, for all radiatively allowed transitions,
together with a set of $N_{\mathrm{levels}}$ equations of the
statistical equilibrium (ESE) \emph{at each optical depth} giving the
populations numbers for all atomic levels.

\section{The multilevel atom case}

Hereafter, we shall describe in details how the GS/SOR numerical
method can be implemented for the case of multilevel atom models in 1D
geometry. For simplicity we shall consider the case of non-overlapping
lines with no background continuum.

\subsection{The MALI method: an overview}

ALI methods have been first generalized to the multilevel atom case by
Rybicky and Hummer~\cite{mali1,mali2}. Their MALI method consists in the
{\em preconditioning} of the equations of statistical equilibrium
(ESE) using an approximate operator $\bar{\Lambda}_{ul}^{*}$ and a
$\bar{J}_{ul}^{\mathrm{eff}}$ mean intensity defined hereafter, and
computed, for all allowed transitions $u \leftrightarrow l$, from
quantitites known from the previous iterative step.

Generally speaking, the line source function for a radiative
transition between the two bound atomic levels $u$ (i.e., the upper
energy level) and $l$ (respectively, the lower one) can be written in
terms of the density of populations $n_{u, \, l}$, the
Einstein~\cite{einstein} coefficients for spontaneous emission
$A_{ul}$, absorption $B_{lu}$ and stimulated emission $B_{ul}$, and
the line absorption $\phi_{\nu}$ and emission $\psi_{\nu}$ profiles
such as

\begin{equation}
S_{ul}{(\nu)} = { {n_{u} A_{ul} { {\psi_{\nu} } } } \over {n_{l}
B_{lu} {{ \phi_{\nu}}} - n_{u} B_{ul} { {\psi_{\nu}}}} } \, .
\label{eq:source}
\end{equation}
Hereafter we shall however restrict our study to complete
redistribution in frequency for which absorption and emission profiles
are identical i.e. $\phi_{\nu} \equiv \psi_{\nu}$ and therefore, the
line source function remains \emph{independent} of the frequency. It
is indeed well-known that such a assumption remains valid for most of
the spectral lines; however, partial redistribution in frequencies
effects should be considered, at the cost of more computational work,
for a proper description of resonance line formation in a diluted
medium (e.g.,~\cite{fp95} and references therein).

Furthermore, leaving out bound-free transitions, at each depth in the
atmosphere, the ESE are usually expressed as a set of
$N_{\mathrm{levels}}$ elementary equations such that
\begin{eqnarray}
  {\sum_{j < i} {[ n_{i} {A_{ij}}
 - (n_{j}
      B_{ji} - n_{i} B_{ij}){\bar{J}_{ij}} ]} } \cr
 -{\sum_{j
      > i} {[ n_{j}  {A_{ji}} - 
 (n_{i} B_{ij}
      - n_{j} B_{ji}){\bar{J}_{ij}} ]} } \cr
 +
{\sum_{j}
    {(n_{i} C_{ij} - n_{j} C_{ji})}} = 0 \, ,
\end{eqnarray}
where the $C_{ij}$ are collisional excitation/deexcitation
rates. Now, introducing the approximate operator into the ESE via the
ALI approximation

\begin{equation}
I_{\nu \Omega} \simeq \Lambda^{*}_{\nu \Omega} [S] +
(\Lambda_{\nu \Omega} - \Lambda^{*}_{\nu \Omega}) [S^{\mathrm{(old)}} ] ~,
\label{eq:approx}
\end{equation}
where $S^{\mathrm{(old)}}$ represents the source function
known from previous iteration and, defining

\begin{equation}
\left\{
\begin{array}{l}
{\bar{J}_{ij} \simeq {\displaystyle \int} 
{\displaystyle {{d \Omega} \over {4 \pi}}}
{\displaystyle \int}
  {\phi_{\nu} \Lambda_{\nu \Omega}^{*} [S] d \nu} +
  \bar{J}^{\rm eff}_{ij} }  \\\\
  \bar{J}^{\rm eff}_{ij} = {\displaystyle \int} 
\displaystyle{ {{d \Omega} \over {4 \pi}} }
{\displaystyle \int} {\phi_{\nu}} {\left( \Lambda_{\nu \Omega}
-  \Lambda_{\nu \Omega}^{*} \right)
[{S^{\mathrm{(old)}}}] d \nu}
\end{array} \, ,
\right.
\end{equation}
where $\bar{J}^{\rm eff}_{ij}$ depends only on known quantities, we
can establish\footnote{Inserting Eq.~(\ref{eq:source}) into
Eq.~(\ref{eq:approx}) and forming the net radiative rate $(n_{j}
B_{ji} - n_{i} B_{ij}){\bar{J}_{ij}}$ in terms of
$\bar{\Lambda}^{*}_{ij}$ and $\bar{J}^{\rm eff}_{ij}$.} after
~\cite{mali1} the following set of preconditioned equations
\begin{eqnarray}
  {\sum_{j < i} {[ n_{i} {A_{ij} (1-\bar{\Lambda}^{*}_{ij})}
 - (n_{j}
      B_{ji} - n_{i} B_{ij}){\bar{J}_{ij}^{\rm eff}} ]} } \cr
 -{\sum_{j
      > i} {[ n_{j}  {A_{ji} (1-\bar{\Lambda}^{*}_{ij})} - 
 (n_{i} B_{ij}
      - n_{j} B_{ji}){\bar{J}_{ij}^{\rm eff}} ]} } \cr
 +
{\sum_{j}
    {(n_{i} C_{ij} - n_{j} C_{ji})}} = 0 \, ,
\label{eq:mali}
\end{eqnarray} 
where 

\begin{equation}
\bar{\Lambda}^{*}_{ij} = \int \displaystyle{{{d \Omega} \over {4 \pi}}}
  \int{\phi_{\nu}\Lambda_{\nu \Omega}^{*} d \nu} \, .
\end{equation}

Heinzel~\cite{ph95} and Paletou~\cite{fp95} showed how to take care
self-consistently of the ionization equilibrium within the MALI
approach by adding a Newton--Raphson scheme to it, for the
determination of the electron density from a set of non-linear ESE
taking into account all kind of bound-free transitions; the same can
be done within the new GS/SOR multilevel numerical schemes.


\subsection{Expliciting GS/SOR with multilevel atoms}

Assume that one has {\em already} swept the grid {\em once}, say from
the illumination-free surface of the atmosphere at $k = 0$, down to
the bottom boundary at $k=\mathrm{ND}$ along direction $- \Omega$. By
analogy with the GS/SOR numerical strategy for the two-level atom
case, in the multilevel atom case we are now going to update all
population numbers at successive depths $k=\mathrm{ND}, ..., 1$
\emph{while sweeping back the grid along the opposite, upward
direction} $+ \Omega$.

Now the population update will be made depth after depth, from the
bottom, up to the surface of the atmosphere by inverting the MALI
preconditioned set of ESE given in Eqs.~(\ref{eq:mali}) \emph{before}
passing to the next depth point. It is a quite straightforward task at the
lower boundary surface since the incident radiation field is known a
priori from the (given) external conditions of illumination.

The situation is however a bit more tricky at the inner grid points.
Once the populations at depth $(k+1)$ have been updated, we shall
advance along direction $+ \Omega$ to the next grid point at depth
$k$. But having changed $\{ n_{j}^{\mathrm{(old)}} \}_{(k+1)}$ to $\{
n_{j}^{\mathrm{(new)}} \}_{(k+1)}$ means that the local absorption
coefficients $\chi_{(k+1)}$ {\em and} the source functions for all
allowed transitions have to be changed accordingly, as well as the
upwind optical depth $\Delta \tau_{(k+1)}^{(\uparrow)}$ along the path
from depth $(k+1)$ to depth $k$. As a consequence, the three
coefficients $\Psi^{(\uparrow)}_{d}$ [$d=(k-1),\, k,\, (k+1)$] used
for the evaluation of the specific intensity along the
short-characteristics are also affected by the local population change
and thus need to be updated. But since $\Delta
\tau_{(k+1)}^{(\uparrow)} = \Delta \tau_{(k-1)}^{(\downarrow)} $
evaluations of the specific intensities made during the first downward
pass must also be corrected accordingly, for consistency.

   \begin{figure}
   \centering
   \includegraphics[width=8cm,angle=-90]{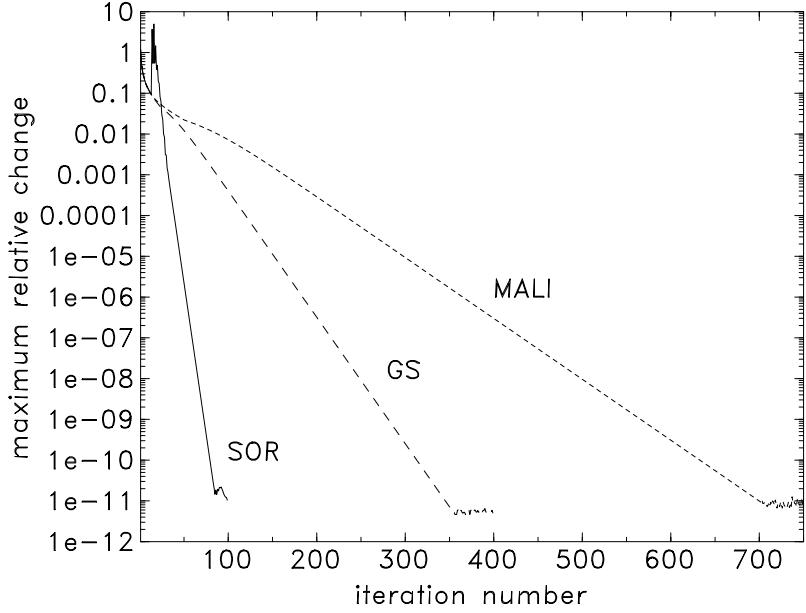}
      \caption{Rates of convergence for the MALI, Gauss-Seidel and SOR
	multilevel iterative processes, respectively. A spatial grid
	of 20 depth points per decade together with a 8 Gauss--Legendre
	angular quadrature and constant Doppler profiles were
	used. The atomic model is a 3-level H\,{\sc i} model atom
	taken from ~\cite{avrett} (see also Table 2). }
         \label{Fig2}
   \end{figure}

So the major point in such implementation is to carefully
``propagate'' the effects of the local population update. Let us, for
instance, consider the specific intensity evaluation at any inner grid
point $k$ in, respectively, direction $- \Omega$ for which we shall
use $(\downarrow)$ superscripts, and $+ \Omega$ (resp. $(\uparrow)$
superscripts); we can therefore write, following Eq.~(\ref{eq:sc})

\begin{equation}
I_{k}^{(\downarrow)} = I_{k-1}^{(\downarrow)}
e^{- \Delta \tau_{(k-1)}^{(\downarrow)}} +
\Psi_{k-1}^{(\downarrow)} S_{k-1}+ \Psi_{k}^{(\downarrow)} S_{k}
+ \Psi_{k+1}^{(\downarrow)} S_{k+1} \, ,
\label{eq:idown}
\end{equation}
while, at the same depth but in the opposite direction, we have

\begin{equation}
I_{k}^{(\uparrow)} = I_{k+1}^{(\uparrow)} e^{- \Delta
\tau_{(k+1)}^{(\uparrow)}} + \Psi_{k+1}^{(\uparrow)} S_{k+1}+
\Psi_{k}^{(\uparrow)} S_{k} + \Psi_{k-1}^{(\uparrow)} S_{k-1} \, .
\label{eq:iup}
\end{equation}
Since the update of the population numbers at depth $(k+1)$ have just
been done by inverting the system of Eqs.~(\ref{eq:mali}), this
generates \emph{for each allowed transition} changes in the
absorption coefficients at line center, now becoming

\begin{equation}
\chi_{(k+1)}^{\mathrm{(new)}} = \left( { {h \nu_{ul} } \over { 4 \pi} }
\right) \left[ n_{l}^{\mathrm{(new)}} B_{lu} - 
n_{u}^{\mathrm{(new)}} B_{ul} \right] \,
\end{equation}
{\em and} in the line source functions, assuming complete
redistribution in frequency, turning into

\begin{equation}
S_{(k+1)}^{(\mathrm{new})} = {{n_{u}^{\mathrm{(new)}} A_{ul}} \over 
{ n_{l}^{\mathrm{(new)}} B_{lu} - n_{u}^{\mathrm{(new)}} B_{ul} }} \, .
\end{equation}
Therefore, one has first to correct $I_{k}^{(\downarrow)}$ for
consistency since previous changes lead, according to
Eq.~(\ref{eq:idown}), to

\begin{equation}
\begin{array}{rl}
I_{k}^{(\downarrow),\,{\mathrm{(new)}}} = &
I_{k-1}^{(\downarrow),\,{\mathrm{(old)}}}
e^{- \Delta \tau_{(k-1)}^{(\downarrow),\,{\mathrm{(old)}}}} +
\Psi_{k-1}^{(\downarrow),\,{\mathrm{(new)}}} S_{k-1}^{{\mathrm{(old)}}} \\\\
+ & 
\Psi_{k}^{(\downarrow),\,{\mathrm{(new)}}} S_{k}^{{\mathrm{(old)}}} + 
\Psi_{k+1}^{(\downarrow),\,{\mathrm{(new)}}} S_{k+1}^{{\mathrm{(new)}}} \,\, ,
\end{array}
\label{eq:corr1}
\end{equation}
which requires that $I_{k-1}^{(\downarrow),\,{\mathrm{(old)}}}$ have
been saved in memory during the first downward pass. This step is
equivalent to the computation of the $\Delta J_{k}^{\mathrm{in}}$
correction pointed out in~\cite{tf1}, Eq.~(39).

   \begin{figure}
   \centering
   \includegraphics[width=8cm,angle=-90]{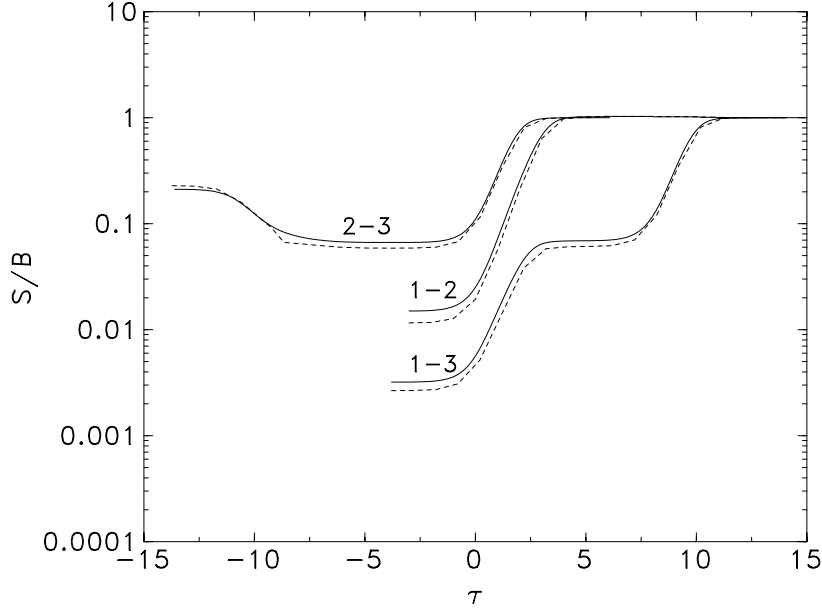}
      \caption{Source functions vs. optical depth for the three
      transitions of our H-model. Solid lines are our GS/SOR
      multilevel results and dashed lines correspond to low-resolution
      results previously published in~\cite{avrett} and~\cite{atanackovic}.}
         \label{Fig3}
   \end{figure}

Then, $I_{k}^{(\uparrow)}$ can be computed in a straightforward manner
via Eq.~(\ref{eq:iup}), which makes it possible to compute
$\bar{\Lambda}_{ij}^{*}$ and $\bar{J}_{ij}^{\mathrm{eff}}$ for all
allowed transitions, then to inject these quantities in the
preconditioned ESE and finally to compute/update locally $\{
n_{j}^{\mathrm{(new)}} \}_{k}$ while in the middle of the upward pass.

However, before advancing to next depth point ($k-1$), it is important
to consider first the various changes induced by the level population
changes respectively on the source functions $S_{k}$, on the optical
depth increments $\Delta \tau_{(k+1)}^{(\uparrow),\,{\mathrm{(new)}}}$
and $\Delta \tau_{(k-1)}^{(\uparrow),\,{\mathrm{(new)}}}$ as well as,
finally, on the three short characteristics coefficients
$\Psi^{(\uparrow)}_{d}$ [$d=(k-1),\, k,\, (k+1)$]. Because of that, we
end up all updates at depth $k$ during the upward pass by the
computation of

\begin{equation}
\begin{array}{rl}
I_{k}^{(\uparrow),\,{\mathrm{(new)}}} = & 
I_{k+1}^{(\uparrow)}
e^{- \Delta \tau_{(k+1)}^{(\uparrow),\,{\mathrm{(new)}}}} +
\Psi_{k+1}^{(\uparrow),\,{\mathrm{(new)}}} S_{k+1}^{{\mathrm{(new)}}} \\\\
 + & 
\Psi_{k}^{(\uparrow),\,{\mathrm{(new)}}} S_{k}^{{\mathrm{(new)}}} + 
\Psi_{k-1}^{(\uparrow),\,{\mathrm{(new)}}} S_{k-1}^{{\mathrm{(old)}}} \,\, .
\end{array}
\label{eq:corr2}
\end{equation}
This last stage is analogous to the correction described by Eq.~(40)
in ~\cite{tf1}.

Finally, a multilevel SOR iterative scheme is built when, at each depth
$k$, all the populations of the excited levels are updated according to

\begin{equation}
n_{k}^{\mathrm{(new)}} = n_{k}^{\mathrm{(old)}} + \omega \Delta
n_{k}^{\mathrm{(GS)}} \, ,
\label{eq:sorpop}
\end{equation}
where $\omega$ is a parameter of the order of 1.5 as discussed below.

\subsection{Numerical recipes}

Several ``recipes'' should be followed when implementing a GS/SOR
solver for multilevel atoms. The first one is to order properly the
various loops. From outer to inner loops one may find indeed: (1) the
directions ($\pm \Omega$) along which the slab will be swept, (2) the
number of allowed radiative transitions, (3) the direction cosines
(i.e., the usual $\mu$'s) and, finally (4) the frequencies.

Upwind and downwind corrections, as described in Eqs.~(\ref{eq:corr1})
and (\ref{eq:corr2}), require some bookkeeping of variables such as all
the $I_{k}^{(\downarrow),\,{\mathrm{(old)}}}$ after the downwind pass
$- \Omega$ for the further computation of the mean intensity entering
the preconditioned ESE i.e., Eqs.~(\ref{eq:mali}).

\begin{table}[h]

\caption{Computation time for the 3-level H model of~\cite{avrett}
obtained with an Intel Pentium-4 clocking @ 3 GHz as a function of the
number of depth points per decade $N_{\tau}$; the stopping criterion
was $R_c = 10^{-10}$ (we also indicate the \# of iterations required
respectively).}

\label{table2}
\centering
\begin{tabular}{c c c c}
\hline
$N_{\tau}$ & MALI & GS & SOR \\
\hline
   5 & 7.644s (160) & 4.595s (78) & 2.325s (39) \\
  10 & 27.852s (326) & 17.148s (161) & 6.096s (55) \\
  15 & 59.768s (483) & 38.997s (239) & 10.203s (63) \\
  20 & 1m43.373s (634) & 1m05.879s (315) & 17.213s (81) \\
  25 & 2m37.637s (780) & 1m40.630s (387) & 26.040s (99) \\
\hline
\end{tabular}
\end{table}

Finally, it is important to realize that the additional computing time
necessary for all the extra-computations required by one GS/SOR
iterative step is very small as compared to the huge saving on the
whole convergence process with respect to MALI (see Table 1).

   \begin{figure}
   \centering
   \includegraphics[width=8cm,angle=-90]{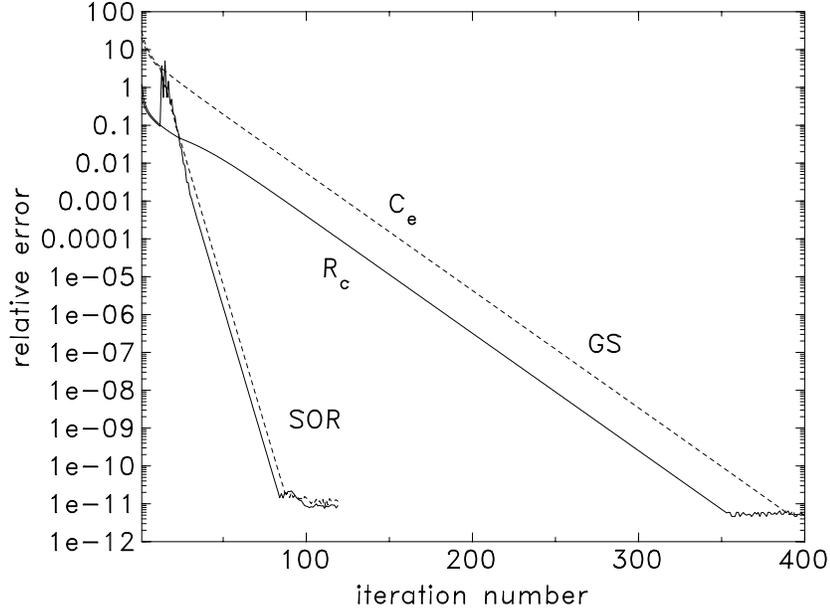}
      \caption{History of the convergence error $C_e$ (dashed
	lines) and of the relative change $R_c$ (full lines) for,
	respectively, the GS and the SOR multilevel iterative
	schemes. Although a small value of $R_c$ does not necessarily
	imply a small enough value of $C_e$ as demonstrated by the GS
	curves, a SOR iterative process including a self-consistent
	evaluation of $\omega$ leads to a scheme such that $R_c \simeq
	C_e$.}
         \label{Fig4}
   \end{figure}

\section{Illustrative examples and discussion}

We have adopted the standard benchmark models for multilevel atom
problems proposed by Avrett~\cite{avrett} and Avrett and
Loeser~\cite{avrettloeser} whose main parameters have also been
summarized in Table 2.

The respective rates of convergence for the MALI, Gauss-Seidel (GS)
and SOR multilevel iterative processes are displayed in
Fig.~\ref{Fig2} where we plotted the maximum relative change on the
level populations from an iteration to another, $R_c$, for the
schematic 3-level H\,{\sc i} model. A spatial grid of 20 depth points
per decade together with a 8 Gauss--Legendre angular quadrature and
constant Doppler profiles were used. It clearly demonstrates how
superior to MALI the multilevel SOR iterative scheme can be for such
refined grids; and it is worth pointing out here again the importance
of grid refinement on the accuracy of ALI-class methods, as
demonstrated by Chevallier et al.~\cite{loic}. Indeed, in such a case
a factor of $\sim 6$ in computing time can be saved, when one iterates
the population numbers down to numerical noise, for instance. We
report a similar behaviour of the iterative processes when dealing
with Avrett's Ca\,{\sc ii} ion model.

\begin{table}
\caption{Input parameters for the H\,{\sc i} and Ca\,{\sc ii}
multilevel benchmark models of Avrett~\cite{avrett}. Statistical weights are
$g_1=2$, $g_2=8$, and $g_3=18$ for H\,{\sc i}, $g_1=2$, $g_2=4$,
$g_3=6$, $g_4=2$, $g_5=4$ for Ca\,{\sc ii}; the temperature of the
atmosphere is 5000 K.}
\label{table1}
\centering
\begin{tabular}{c c c c c c}
\hline
element & $u$ & $l$ & $A_{ul}$ & $C_{ul}$ & $\nu_{ul}$ \\
\hline
   H & 2 & 1 & $4.68 \times 10^{8}$ & $10^{5}$ & $2.47 \times 10^{15}$ \\
   H & 3 & 1 & $5.54 \times 10^{7}$ & $10^{5}$ & $2.93 \times 10^{15}$ \\
   H & 3 & 2 & $4.39 \times 10^{7}$ & $10^{5}$ & -- \\ \hline
  Ca & 2 & 1 & forbidden & $8.2 \times 10^{3}$ & $4.10 \times 10^{14}$ \\
  Ca & 3 & 1 & forbidden & $8.2 \times 10^{3}$ & $4.12 \times 10^{14}$ \\
  Ca & 4 & 1 & $1.4 \times 10^{8}$ & $5.1\times10^{4}$ & $7.56 \times 10^{14}$ \\
  Ca & 5 & 1 & $1.4 \times 10^{8}$ & $5.1\times 10^{4}$ & $7.63 \times 10^{14}$ \\
  Ca & 3 & 2 & forbidden & $10^{7}$ & -- \\
  Ca & 4 & 2 & $7.8 \times 10^{6}$ & $1.6 \times 10^{5}$ & -- \\
  Ca & 5 & 2 & $8.1 \times 10^{5}$ & $1.6 \times 10^{4}$ & -- \\
  Ca & 4 & 3 & forbidden & $10^{3}$ & -- \\
  Ca & 5 & 3 & $7.2 \times 10^{6}$ & $1.4 \times 10^{5}$ & -- \\
  Ca & 5 & 4 & forbidden & $4.8 \times 10^{6}$ & -- \\
\hline
\end{tabular}
\end{table}

In Fig.~\ref{Fig3}, we display (solid lines) the source function
vs. line optical depth for the 3 transitions allowed by our schematic
H-model. Dashed lines correspond to the old benchmark results
of~\cite{avrett}; differences are coming not only from very different
numerical schemes but also from large differences in the angular and
frequency quadratures. Our results are also in good agreement with
those of~\cite{atanackovic} and may serve, given the high-level of
grid refinement we adopted, as new benchmark results for multilevel atom
cases.

The main potential drawback of SOR methods is that it relies on the
choice of a relaxation parameter $\omega$ whose optimal value is a
priori unknown. However, it was proposed~\cite{tf1} a quite robust
numerical procedure in order to estimate \emph{self-consistently} a
close-to-optimal $\omega$ after having run a few ``pure GS''
iterations. We followed their recommendations and, indeed, found a
posteriori values close to ``optimal'' ones deduced from experimental
runs using a prescribed $\omega$.

Finally, in Fig.~\ref{Fig4}, we plotted the history of the convergence
error $C_e$ defined as

\begin{equation}
C_{e} = \mathrm{max} \left( { {\mid n(itr) - n(\infty) \mid} \over
{n(\infty)} }
\right) \, ,
\end{equation}
following ~\cite{lhaft}, where $itr$ is the iteration number and
$n(\infty)$ is the fully converged solution, and of the relative
change $R_c$ for, respectively, the GS and the SOR multilevel
iterative schemes.  It is important to note again that reaching a
small value of $R_c$, which is indeed the most direct control
parameter of the iterative process, does \emph{not} necessarily imply
a small enough value of $C_e$ to guarantee convergence; and this is
shown by the GS curves. However, an optimal SOR iterative process
including the self-consistent evaluation of $\omega$ as proposed
in~\cite{tf1}, leads to a better-controlled process since $R_c$ is
just slightly lower than $C_e$, as shown in Fig.~\ref{Fig4}.

The issue of having a reliable stopping criterion is of course
critical to any iterative method and, for the specific case of
numerical radiative transfer, Auer et al.~\cite{lhaft} and Fabiani Bendicho
et al.~\cite{fta2} addressed it successfully by adopting \emph{multi-grid}
methods. We finally refer the reader to this later work where it is
demonstrated how the \emph{combination} of GS/SOR iterative schemes
together with multi-grid techniques lead to extremely powerful
techniques for the solution of complex radiative transfer problems.

\section*{Acknowledgements}
Our thanks go to Drs Lo\"{\i}c Chevallier and Fran\c cois Rincon for their
careful reading of an early version of this paper.  All figures have
been made using the freeware package ANA ({\tt http://ana.lmsal.com/})
developed at the Lockheed Martin Solar and Astrophysics Laboratory by
Dick Shine and Louis Strous.


\begin{thebibliography}{00}

\bibitem{tf1} Trujillo Bueno J, Fabiani Bendicho P. A novel iterative
scheme for the very fast and accurate solution of non-LTE radiative
transfer problems. Astrophys. J. 1995;455:646

\bibitem{lhafp} Auer LH, Paletou F.  Two-dimensional radiative
transfer with partial frequency redistribution I. General
method. Astron. Astrophys.  1994;285:675

\bibitem{lhaft} Auer LH, Fabiani Bendicho P, Trujillo Bueno J.
Multidimensional radiative transfer with multilevel atoms. I. ALI
method with preconditioning of the rate
equations. Astron. Astrophys. 1994;292:599

\bibitem{fta2} Fabiani Bendicho P, Trujillo Bueno J, Auer
LH. Multidimensional radiative transfer with multilevel atoms. II. The
non-linear multigrid method. Astron. Astrophys. 1997;324:161

\bibitem{vanoort} van Noort M, Hubeny I, Lanz T. Multidimensional
Non-LTE radiative transfer. I. A universal two-dimensional
short-characteristics scheme for cartesian, spherical, and cylindrical
coordinate systems. Astrophys. J. 2002;568:1066

\bibitem{goutte} Gouttebroze P. Radiative transfer in cylindrical
threads with incident radiation. Astron. Astrophys. 2004;413:733

\bibitem{tm} Trujillo Bueno J, Manso Sainz R. Iterative methods for
the non-LTE transfer of polarized radiation: resonance line
polarization in one-dimensional
atmospheres. Astrophys. J. 1999;516:436

\bibitem{javier} Trujillo Bueno J. In Hubeny I, Mihalas D, Werner K,
  editors. ASP Conf. Ser. 288, Stellar Atmosphere Modeling,
  Astronomical Society of the Pacific, San Francisco, 2003. p. 551

\bibitem{ivan} Hubeny I. In Hubeny I, Mihalas D, Werner K,
  editors. ASP Conf. Ser. 288, Stellar Atmosphere Modeling,
  Astronomical Society of the Pacific, San Francisco, 2003. p. 17

\bibitem{oab} Olson GL, Auer LH, Buchler JR. A rapidly convergent
iterative solution of the non-LTE line radiation transfer problem.
JQSRT 1986;35:431

\bibitem{mihalas} Mihalas D, Stellar Atmospheres.  Freeman, San
  Francisco, 1978

\bibitem{rob} Rutten RJ. In Hubeny I, Mihalas D, Werner K,
  editors. ASP Conf. Ser. 288, Stellar Atmosphere Modeling,
  Astronomical Society of the Pacific, San Francisco, 2003. p. 99 (see
  also Radiative Transfer in Stellar Atmospheres, Sterrekunding
  Instituut Utrecht\footnote{{\tt
  http://www.fys.ruu.nl/$\sim$rutten/Astronomy\_course.html}})

\bibitem{young} Young DM. Iterative Solution of Large Linear
  Systems. Academic Press, New York, 1971

\bibitem{hackbusch} Hackbusch W. Multi-Grid Methods and
  Applications. Springer--Verlag, Berlin, 1985

\bibitem{briggs} Briggs WL, Henson VE, McCormick SF.  A Multigrid
  Tutorial. Soc. for Industrial and Applied Math., Philadelphia, 2000

\bibitem{mali1} Rybicki GB, Hummer DG. An accelerated lambda iteration
method for multilevel radiative transfer. I. Non-overlapping lines
with background continuum. Astron. Astrophys. 1991;245:171

\bibitem{sc1} Olson GL, Kunasz PB. Short characteristic solution of
the non-LTE line transfer problem by operator perturbation. I. The
one-dimensional planar slab. JQSRT 1987;38:325

\bibitem{sc2} Kunasz PB, Auer LH.  Short characteristic integration of
radiative transfer problems - Formal solution in two-dimensional
slabs. JQSRT 1988;39:67

\bibitem{mali2} Rybicki GB, Hummer DG. An accelerated lambda iteration
method for multilevel radiative transfer. II. Overlapping transitions
with full continuum 1992, Astron. Astrophys. 1992;262:209

\bibitem{einstein} Einstein A. Zur quantentheorie der
strahlung. Phys. Z. 1917;18:121-128

\bibitem{ph95} Heinzel P. Multilevel NLTE radiative transfer in
isolated atmospheric structures: implementation of the
MALI-technique. Astron. Astrophys. 1995;299:563

\bibitem{fp95} Paletou F. Two-dimensional multilevel radiative
transfer with standard partial frequency redistribution in isolated
solar atmospheric structures. Astron. Astrophys. 1995;302:587

%

\bibitem{avrett} Avrett EH. in Athay RG, Mathis J, Skumanich A, editors.
  Resonance Lines in Astrophysics, National Center for Atmospheric
  Research, Boulder, 1968, 27

\bibitem{avrettloeser} Avrett EH, Loeser R. in Kalkofen W, editor.
  Numerical Radiative Transfer, Cambridge Univ.  Press, Cambridge,
  1987, 135

\bibitem{loic} Chevallier L, Paletou F, Rutily B. On the accuracy of
the ALI method for solving the radiative transfer
equation. Astron. Astrophys.  2003;411:221

\bibitem{atanackovic} Atanackovi\'c-Vukmanovi\'c O, Crivellari L,
Simonneau E. A forth-and-back implicit
$\Lambda$-iteration. Astrophys. J. 1997;487:735

\end{thebibliography}
\end{document}